\documentclass [final,onecolumn,notitlepage,a4paper,12pt]{article}
\usepackage [a4paper] {geometry}
\usepackage{amssymb}
\usepackage{amsfonts}
\usepackage{amsmath}
\usepackage{hyperref}
\usepackage{bbm}
\usepackage{multirow}

\numberwithin{equation}{section}
\newcommand{\beq}{\begin{eqnarray}}
\newcommand{\eeq}{\end{eqnarray}}
\newcommand{\bsp}{\begin{split}}
\newcommand{\esp}{\end{split}}
\newcommand{\ds}{\slash\!\!\!\!\!\:\:\partial}

\newcommand{\lsim}{\mathrel{\rlap{\lower4pt\hbox{\hskip1pt$\sim$}}
    \raise1pt\hbox{$<$}}}         %less than or approx. symbol
\newcommand{\gsim}{\mathrel{\rlap{\lower4pt\hbox{\hskip1pt$\sim$}}
    \raise1pt\hbox{$>$}}}         %greater than or approx. symbol

\begin{document}
%\maketitle

%%%%%%%%%%%%%%%%%%%%%%%%%%%%%%%%%%%%%%%%%%%%%%%%%%%%%%%%%%%%%%%%%%%%

\begin{titlepage}
\begin{center}

%\hfill xxx\\

\rightline{\small WIS/16/08-JUL-DPP}
\vskip 2cm

{\Large \bf  Splitting the Wino Multiplet by Higher-Dimensional Operators in Anomaly Mediation}

\vskip 1.2cm

{\bf Micha Berkooz and Yonit Hochberg }
\
\vskip 0.8cm
{\em Department of Particle Physics,\\ Weizmann Institute of Science,
Rehovot 76100, Israel}  \\
{\tt Micha.Berkooz@weizmann.ac.il, Yonit.Hochberg@weizmann.ac.il}
\vskip 0.4cm

\end{center}

\vskip 1cm

\begin{center} {\bf ABSTRACT } \end{center}
%\vspace{-2mm}

\noindent

In a class of AMSB models, the splitting in the Wino multiplet turns out to be very small, such as the often-quoted 170 MeV in minimal AMSB, which originates from MSSM loops.  Such a small mass gap is potentially a window into higher scale physics, as it may be sensitive to higher-dimensional operators.  We show that still within AMSB one can get a much larger splitting in the Wino multiplet---a few GeV---if the scale of the new physics is comparable to the gravitino mass (which is indeed often the scale of new physics in anomaly mediation). %This is indeed often the scale of new physics in anomaly mediation.

\vfill

%\today
%\noindent

\end{titlepage}

%%%%%%%%%%%%%%%%%%%%%%%%%%%%%%%%%%%%%%%%%%%%%%%%%%%%%%%%%%%%%%%%

\section{Introduction}\label{sec:intro}

Supersymmetry (SUSY) is a suggested symmetry of nature, which, if it
exists, must be broken.  An attractive idea for communicating SUSY breaking to the Supersymmetric Standard Model (SSM) is the notion of anomaly mediation - AMSB
\cite{Randall:1998uk, Giudice:1998xp}. AMSB predicts all soft
SUSY-breaking parameters in terms of the gravitino mass and the
$\beta$ and $\gamma$ functions of the low energy theory.

The gaugino masses are proportional to the gauge $\beta$ functions,
while the scalar masses and soft trilinear couplings are
proportional to the anomalous dimensions of the corresponding scalar
fields.   In pure AMSB, the mass spectrum is such that the neutral
Wino is the LSP, with a nearly degenerate charged Wino as the NLSP
\cite{Feng:1999fu, Gherghetta:1999sw}. The dominant contribution to
the mass splitting of the Wino multiplet comes from one-loop
corrections, mainly from gauge loops. A severe problem of
the AMSB scenario is the tachyonic slepton spectrum, an issue which
has been addressed in various ways
\cite{Randall:1998uk}, \cite{Gherghetta:1999sw}-\cite{Kikuchi:2008gj}.
%\cite{Dermisek:2007qi},
The resulting spectrum predictions may vary, depending on
the details of the resolution.%A severe problem of
%the AMSB scenario is the tachyonic slepton spectrum, an issue which
%has been addressed in various ways
%\cite{Randall:1998uk}, \cite{Gherghetta:1999sw}-\cite{Kikuchi:2008gj},
%%\cite{Dermisek:2007qi},
%with some
%resolutions disrupting UV insensitivity and others leaving it
%intact.  The resulting spectrum predictions may vary, depending on
%the details of the resolution.

We will use the terminology {\it standard anomaly mediation} to
refer to the following scenario:
\begin{itemize}
\item The tachyonic slepton spectrum has been cured in some manner,
leaving the gaugino mass predictions of AMSB intact (at least to
leading order).
\item The LSP and NLSP are Wino-like and are highly
degenerate, with mass splitting $\sim 150-300$ MeV.
\end{itemize}
This is the characteristic phenomenology of anomaly mediation.  Such
a scenario is realized, for instance, in the majority of parameter
space in the phenomenological approach of minimal anomaly mediation
(mAMSB) \cite{Gherghetta:1999sw, Feng:1999hg}, where the AMSB scalar
masses squared are enhanced by an additional universal
contribution of $m_0^2$, such that the sleptons are non-tachyonic and above LEP bounds.
The standard AMSB
scenario is, however, broader than mAMSB.

In this work, we consider adding higher-dimensional operators to the
Lagrangian within the anomaly mediation framework. We will restrict ourselves to operators arising via a threshold, which we denote by $\Lambda$, that exists and is non-zero in the $m_{3/2}\rightarrow0$ limit, in which SUSY is broken only by AMSB.  We refer to such a threshold as a supersymmetric one, by which we mean it exists in the SUSY limit, though is not SUSY preserving.

Given that AMSB determines the mass of the gauginos in terms of
the low energy couplings, there is no useful notion of changing the
term $\int d^4x \tilde{W}\tilde{W}$ directly.  However, in the
presence of higher-dimensional operators, additional mixing terms
can appear, changing the fermion masses after
electroweak breaking and once the new mass matrices are diagonalized.
These are the operators which we will use.  The effects which we discuss are therefore analytic negative powers of $\Lambda$ which are activated by VEVs of the Higgs.  Indeed they usually play no role in AMSB, except that here we are discussing their effect on an unusually small number---the splitting in the Wino multiplet.

%It will be interesting to extend this to cases where this SUSY threshold is zero (and becomes an effective threshold only after SUSY breaking) such as in \cite{Pomarol:1999ie, Nelson:2002sa, Hsieh:2006ig}, or to more general mixed anomaly- and gauge-mediation scenarios, but we will not do so here. The advantage of
%this low energy effective action approach is that it captures the
%bounds and dynamics that are associated with any higher energy
%scale, as long as the condition above is met.

Within this framework we already get that higher-dimensional operators, suppressed by $\Lambda$, lead to a significant splitting in the Winos. The interesting effects arise primarily when the scale of new physics is of order the gravitino mass. In AMSB, indeed one often finds new physics solving the tachyonic slepton problem at such a scale.  We will assume throughout the paper that such new physics directly affects only the scalar sector and not the Winos and Bino (at least to leading order).

In this regime of $m_{3/2}\lesssim\Lambda$, the lightest neutralino and chargino are predominantly Wino-like, composing the LSP and NLSP. At
leading order the behavior is thus similar to the standard anomaly
mediation scenario.  Beyond leading order\footnote{In addition, smaller values of $m_{3/2}$ than are usually allowed are
now accessible for moderate values of $\tan
\beta$, since the LSP mass can be enhanced even for low $m_{3/2}$
values.} we find that large tree level mass
splitting between the two can be obtained, of up to a few GeV (and even higher), much larger
than the often-quoted (loop) number of 170 MeV.

We will consider operators resulting from new physics, but will restrict ourselves only to a qualitative dimensional analysis.  In particular, when the SUSY breaking of the threshold becomes large, there are quantitative corrections that should be taken into account, however they do not change the qualitative results.  %Similarly, we will restrict ourselves to operators that do not contain covariant derivatives of the compensator.
 %
% but not containing covariant derivatives of the compensator.  Such analysis holds for thresholds that are nearly supersymmetric; when the SUSY-breaking of a threshold becomes large, there are quantitative corrections that should be taken into account.  In this paper, however, we are interested only in a qualitative discussion, which is captured even without considering such corrections.

As we focus mainly on the dimensional analysis, we will not keep track of the dependence on the gauge couplings.  In principle, the higher-dimensional operators which we use could have additional coupling factors, but this could be compensated for instance by having the threshold sector be large enough.  In order to insert the coupling dependence and the degeneracy of the threshold, one needs to scale our results appropriately.

In addition to requiring $m_{3/2}<\Lambda$, we consider another source of constraints on the
coefficients of these operators, i.e., on the values of $\Lambda$ and $m_{3/2}$.
These operators, or more precisely their SUSY partners, introduce
additional terms into the Lagrangian, among which we find mixing
terms between the weak and hypercharge gauge field-strengths. The
operators are thus restricted by the electroweak precision variable
$S$ already at tree level. This constraint, however, will turn out not to be the main constraint (once we impose that $m_{3/2}$ is large enough to avoid very light Winos). Rather, the main constraint will simply be the requirement $m_{3/2}<\Lambda$, which we will assume in order to ensure stability of threshold.

Before proceeding, one should reiterate that we are focusing on the
spectrum in the neutralino and chargino sectors, whereas determining
the complete SUSY partner spectrum entails of course assumptions in
the scalar sector beyond AMSB.  Of course, the discussed mass modifications in
the neutralino and chargino sectors still hold regardless of the
scalar spectrum; the resulting fermion sparticle spectrum may simply
not contain the two lightest sparticles, although we will still refer to them throughout the paper as the LSP and NLSP.

The outline of this paper is the following. In Section \ref{sec:renorm} we discuss the renormalizable terms related to the neutralino and chargino sectors, as well as the electroweak $S$ parameter.  In Section \ref{sec:non renorm} we explain the rules regarding the use of higher-dimensional operators. Section \ref{sec:almost} discusses the phenomenology of the region $m_{3/ 2}\lesssim\Lambda$. In Appendix \ref{app:higher} we address the inclusion of some additional terms when $m_{3/2}\sim \Lambda$.  Appendix \ref{app:almost origins} contains a discussion on the origin of orders of magnitude for the dominant non-renormalizable terms. A summary is already contained in
the introduction and hence is omitted.

%%%%%%%%%%%%%%%%%%%%%%%%%%%%%%%%%%%%%%%%%%%%%%%%%%%%%%%%%%%%%%%%%%%%%

\section{Renormalizable terms and electroweak sector bounds}
\label{sec:renorm}

%%%%%%%%%%%%%%%%%%%%%%%%%%%%%%%%%%%%%%%%%%

\subsection{Some notation}\label{notation}
\label{subsec:not}

We start by introducing some notation.  We consider contributions to
the following Lagrangian terms:
\begin{itemize}

\item
  \textbf{Gauge field strength mixing:}\\
  $B_{\mu \nu}$ ($W_{\mu \nu}^a$) denotes the field-strength tensor for the hypercharge $U(1)_Y$ (weak $SU(2)_W$) gauge group. The electromagnetically neutral ones are packaged in $G_{\mu \nu}=\left( W_{\mu \nu}^3, B_{\mu \nu}  \right)^T$.  The relevant terms in the Lagrangian are
  \begin{eqnarray}
  \label{defcls}
    \begin{split}
      G_{\mu \nu}^T {\cal{S}} G^{\mu \nu},
    \end{split}
  \end{eqnarray}
  where in the MSSM, the matrix ${\cal{S}}$ is simply ${\cal{S}}=-\frac{1}{4}\mathbbm{1_{2\times 2}}$.  %\left(\begin{array}{cc}-\frac{1}{4} & 0 \\ 0 & -\frac{1}{4}  \end{array}\right)$.

  \item
  \textbf{Fermionic kinetic mixing:}\\
  In the basis $
  \psi=\left(-i\tilde{B}, -i\tilde{W^3}, \tilde{H_d}^0,\tilde{H_u}^0  \right)^T$, we denote the fermionic kinetic mixing by $-i\psi^T K \ds \bar{\psi}$.
  In the MSSM, $K$ is simply the $4\times4$ unit matrix. This matrix will not play a significant role below.

  \item
  \textbf{Neutralino mass terms:}\\
  The neutralino mass terms of the MSSM are $-\frac{1}{2}(\psi)^T M_{\tilde{N}} \psi+c.c.$, where
\begin{equation}
\begin{split}
  M_{\tilde{N}}=\mbox{\fontsize{10}{14}\selectfont $\left(\begin{array}{cccc}M_1 & 0 & -m_Z c_\beta s_W & m_Z s_\beta s_W \\ 0 & M_2 &m_Z c_\beta c_W & -m_Z s_\beta c_W \\
    -m_Z c_ \beta s_W & m_Z c_\beta c_W & 0 & -\mu \\ m_Z s_\beta s_W & -m_Z s_\beta c_W & -\mu &0  \end{array}\right)
    \equiv\left(\begin{array}{cccc}M_1 & 0 & A_1 & A_2 \\ 0 & M_2 &A_3 & A_4 \\
    A_1 & A_3 & 0 & -\mu \\ A_2 & A_4 & -\mu &0  \end{array}\right)$}
    \end{split}
\end{equation}
and $s_\beta \equiv \sin \beta, c_\beta \equiv \cos \beta,s_W
\equiv \sin \theta_W,c_W \equiv \cos\theta_W$.

\item
\textbf{Chargino mass terms:}\\
The chargino mass terms are $(\psi^-)^T M_{\tilde{C}} \psi^+ +h.c.$, where $\psi^{\pm}=(-i\tilde{W}^{\pm},\tilde{H}^\pm)$ and
\begin{eqnarray}
  \begin{split}
    M_{\tilde{C}}=\left(\begin{array}{cc}
M_2 & \sqrt{2}m_W s_\beta \\ \sqrt{2}m_W c_\beta & \mu
\end{array} \right)\equiv \left(\begin{array}{cc}
M_2 & -\sqrt{2}A_4 \\ \sqrt{2}A_3 & \mu
\end{array} \right)\:.
  \end{split}
\end{eqnarray}

\end{itemize}

%%%%%%%%%%%%%%%%%%%%%%%%%%%%%%%%%%%%%%%%%%%%%%%%%%%%%%%

\subsection{Orders of magnitude of renormalizable terms}
\label{subsec:renorm size}

In the anomaly mediation scenario with renormalizable terms only, the above Lagrangian terms are as in the MSSM, with anomaly mediation dictating the gaugino masses $M_1$ and $M_2$.  These are proportional to the $\beta$ functions, and are given at the weak scale $m_Z$ by (e.g., \cite{Giudice:1998xp})
\begin{eqnarray}
\begin{split}
&M_1=\frac{11 \alpha_w}{4 \pi \cos^2 \theta_W}m_{3/2}=8.9\times10^{-3} m_{3/2}\;,\\
&M_2=\frac{\alpha_w}{4 \pi \sin^2 \theta_W}m_{3/2}=2.7\times 10^{-3}
m_{3/2}\;.
\end{split}
\end{eqnarray}
where %$\alpha_w(m_Z)\sim1/128$ and $\sin ^2\theta_W(m_Z)\sim0.23$.
\begin{equation}
\alpha_w(m_Z)\sim\frac{1}{128}\;,\ \ \sin ^2\theta_W(m_Z)\sim0.23\;.
\end{equation}
We assume a $\mu$ term consistent with successful electroweak symmetry breaking has been generated, and is of order a few hundred GeV.  The electroweak mass mixing terms are given by
\begin{eqnarray}\label{eq:Ai}
  \begin{split}
    A_1&=-g' \frac{v_d}{\sqrt{2}}\sim -44 \cos \beta\
    \mathrm{GeV},\ \
    A_2= g' \frac{v_u}{\sqrt{2}}\sim 44 \sin \beta\ \mathrm{GeV},\\
    A_3&= g \frac{v_d}{\sqrt{2}} \sim 80 \cos \beta\ \mathrm{GeV},\ \
    A_4=-g \frac{v_u}{\sqrt{2}}\sim -80 \sin \beta\ \mathrm{GeV}\;.
  \end{split}
\end{eqnarray}

%%%%%%%%%%%%%%%%%%%%%%%%%%%%%%%%%%%%%%%%%%

\subsection{Electroweak $S$ and $T$ bounds}
\label{subsec:S}

The operators that induce mixing in the neutralino and chargino sectors have SUSY
partners that affect the Yang-Mills terms in the Lagrangian.
They will therefore be constrained by electroweak data, particularly from the precision electroweak variables $S$ and $T$
\cite{Peskin:1991sw, Grinstein:1991cd}. Recall that the $S,T,U$
variables parameterize the oblique corrections from new physics at
high energy.  $S$ is a measure of the total size of the new sector,
while $T$ is a measure of the total weak-isospin breaking induced by
it.  These are effects of vacuum polarization diagrams and are not
corrections from the Standard Model (SM).  The parameters can be defined
by \cite{Lim:2007ea}
\begin{eqnarray}
  \begin{split}
    S=-\frac{16 \pi}{g g'} \Pi' _{3Y}\;,\ \  T=-\frac{4 \pi}{g^2 \sin \theta_W^2}\frac{\Delta M^2}{M_W^2}\;,
  \end{split}
\end{eqnarray}
where $\Pi_{3Y}'\equiv \frac{d^2}{dp^2}\Pi_{3Y}(p^2)|_{p^2=0}$, $\Pi_{3Y}$ denotes the $g_{\mu \nu}$ part of the gauge-boson self energy between the neutral weak and hypercharge gauge bosons, and  $\Delta M^2 \equiv \delta M_{W^3}^2-\delta M_{W^\pm}^2$, with $\delta M_{W^3}^2,\ \delta M_{W^\pm}^2$ denoting corrections to the neutral and charged $SU(2)_W$ gauge-bosons mass-squared.

The experimental values of the $S$ and $T$ parameters are \cite{Amsler:2008zz}
\begin{eqnarray}
      S=-0.10\pm 0.10\;, \ \   T=-0.08\pm 0.11
      \label{ST values}
\end{eqnarray}
for $m_h=117$ GeV, and $S$ decreases monotonically as $m_h$ increases.

Now, assume that we have a term of the form $\eta W_{\mu \nu}^3
B^{\mu \nu}$ in the Lagrangian.  We can translate such a term into
\begin{eqnarray}
  \begin{split}
    \eta W_{\mu \nu}^3 B^{\mu \nu}&=2\eta(p^2 g_{\mu \nu}-p_\mu p_\nu)W^{3 \mu} B^\nu
  \end{split}
\end{eqnarray}
in momentum space.  By definition, this contributes to the $S$ parameter by $S=-32 \pi \eta/g g'$,
appearing in the off-diagonal entries of the matrix
${\cal{S}}$ in \eqref{defcls} as
\begin{eqnarray}
  {\cal S}_{\mathrm{off-diag}}=\frac{\eta}{2}=-\frac{g g' S}{64 \pi}\sim -S\times 10^{-3}\;.
  \label{S ref}
\end{eqnarray}
Enforcing the bound on $S$ thus constrains $\eta$.  For our
purposes, this is a constraint on the scales $\Lambda$ and
$m_{3/2}$.

In principle, we can have terms in the Lagrangian affecting the $T$
parameter as well.  If there are no tree level contributions to $T$,
loop corrections inducing mass splitting between the $W$ bosons are
important.  However, in the event that $S$ has tree level
contributions while $T$ does not, the constraint on the scale
$\Lambda$ coming from $S$ will be stronger than that coming from
$T$.  This is a simple consequence of the following.  The contribution to $S$
coming from tree level gives $S={\cal{O}}(\mathrm{tree \
level})/g g'$, while the contribution to $T$ coming from
one-loop diagrams gives $T={\cal{O}}(\mathrm{tree\ level})$.  So,
roughly speaking, $T\sim g g' S$.  Since $T^{\mathrm{th}}<S^{\mathrm{th}}$ and (\ref{ST values}) gives $\Delta T^{\mathrm{exp}} \sim \Delta S^{\mathrm{exp}}$, the stronger constraint comes from the $S$ parameter.  Since we are considering higher-dimensional operators contributing to the fermionic SUSY partners, splitting in the weak gauge boson sector will occur only at loop level, and so we will be concerned with the constraint from $S$ alone.

%%%%%%%%%%%%%%%%%%%%%%%%%%%%%%%%%%%%%%%%%%%%%%%%%%%%%

\section{Non-renormalizable corrections}
\label{sec:non renorm}

%\subsection{The use of higher-dimensional operators}
We are interested in how non-renormlizable terms in the low energy effective action change one of AMSB's canonical predictions---the splitting in the Wino multiplet when it constitutes the lightest set of SUSY partners.  We work in a two-step procedure.  In the first step we assume that, in the absence of SUSY breaking, the low energy theory below a supersymmetric scale $\Lambda$ is the MSSM.  In the next step we break SUSY using the compensator field in this low energy theory.  Throughout we will assume that SUSY is broken only by AMSB.

We will not specify the details of the sector at $\Lambda$, and will only be interested in the operators it induces on the MSSM.  We assume that the threshold exists and is non-zero in the $m_{3/2}\rightarrow0$ limit.\footnote{Thus, this class of models does not include the analyses of \cite{Pomarol:1999ie, Nelson:2002sa, Hsieh:2006ig}, which are effectively a mixture of anomaly- and gauge-mediation SUSY breaking.  The analysis here focuses on the cases with a dominant contribution of the former.}  We assume the spectrum of gauginos is similar to that of standard AMSB, and the main effect that we focus on is the splitting in the Wino multiplet.

The rules that we will use are therefore the following:

\begin{itemize}

\item Writing down the SUSY Lagrangian below $\Lambda$: We consider higher-dimensional operators contributing to the
neutralino and chargino sectors, involving Higgs VEVs. All operators are suppressed by the same scale $\Lambda$, which is
above the weak scale. Each non-renormalizable contribution comes with a general coefficient ${\cal{O}}(1)$.

\item
Using the compensator $\phi$ with background value $\phi=1+\theta^2
m_{3/2}$, the Lagrangian which we use is \cite{Pomarol:1999ie}:
\begin{eqnarray}
  \begin{split}
    \int d^4 \theta \phi \phi^\dagger{\cal{K}}\left(\frac{\phi^{1/2}}{\phi^\dagger}D_\alpha, \frac{Q}{\phi}, \frac{W_\alpha}{\phi^{3/2}}, V\right)+\left(\int d^2 \theta \phi^3 {\cal{W}}\left(\frac{Q}{\phi}, \frac{W_\alpha}{\phi^{3/2}}\right)+h.c.\right),
  \end{split}
  \label{lagrule}
\end{eqnarray}
where for simplicity the dependence of ${\cal{K}}$ on $D_{\dot{\alpha}}^\dagger$, $Q^\dagger$, etc. has been omitted.
The use of the conformal compensator formalism is a matter of
convenience, and dictates the form of the non-renormalizable
operators.  AMSB-like contributions exist independently from any
particular formalism though \cite{Dine:2007me} (and see \cite{deAlwis:2008aq}).

\end{itemize}

The AMSB predictions deduced from this formalism capture the leading order behavior in small SUSY-breaking $m_{3/2}$.  For a qualitative discussion, this leading order prescription will suffice. In Section \ref{sec:almost} we discuss this range, but push $m_{3/2}$ towards $\Lambda$---we can do so since we are performing dimensional analysis---and in this case the only hierarchy that remains is that of $v/\Lambda$ (which is the same as $v/m_{3/2}$), where $v$ is the VEV of the Higgs. We should emphasize though that we will still restrict ourselves, even when $m_{3/2}$ is taken not much below $\Lambda$, to a Lagrangian in which only $\phi$ and the MSSM fields appear.

 One could be concerned that at $m_{3/2}\sim\Lambda$, higher powers in $m_{3/2}/\Lambda$ should be kept, since they could be kicking in when the two scales are becoming equal.  However, the effects we are taking into account are the leading order ones, and any other contributions will vary the results by ${\cal{O}}(1)$, which we will not be concerned with here.  As a partial check, in Appendix \ref{app:higher} we consider terms which have the highest power of $m_{3/2}/\Lambda$ but are still in the Lagrangian \eqref{lagrule}. This will introduce a new host of operators which kick in at $m_{3/2}\sim \Lambda$. However, by examining these operators, we show that they do not affect the qualitative behavior.

We can now proceed and evaluate the order of magnitude of the
contributions of the non-renormalizable operators in the Lagrangian \eqref{lagrule} to the various
Lagrangian terms of Section \ref{notation}. An operator of dimension
$D$ can appear in the Lagrangian as $O_D/\Lambda^{D-4}$. However, in
the presence of SUSY breaking, operators can appear with additional
powers of $m_{3/2}/\Lambda$---the pattern of which is rather
restricted in AMSB (under the discussed assumptions).  If the operator comes
from an $\int d^2\theta$ $F$-term, then the compensator field $\phi$
appears holomorphically, and so, in addition to SUSY-preserving
terms, we will get contributions with one power of $m_{3/2}/\Lambda$.  K\"ahler $\int d^4 \theta$ terms can give us either SUSY-preserving or SUSY-breaking terms, with up to two powers of $m_{3/2}/\Lambda$.  The decoupling limit---the limit in which we return to the
standard predictions of AMSB---is given by
$\Lambda\rightarrow\infty$, keeping $m_{3/2}$ and all SM quantities fixed.

We focus on $m_{3/2}<\Lambda$.  For $\tan\beta=1$, the coefficient of a higher-dimensional operator will be dominated by the lowest possible power of $m_{3/2}/\Lambda$ for this operator.  For $\tan \beta>1$, the ratio between $v_u$ and
$v_d$ slightly alters this naive classification---contributions containing $v_d$
factors and lowest powers of $m_{3/2}$ must be compared to
contributions containing $v_u$ and higher powers of
$m_{3/2}$.

%%%%%%%%%%%%%%%%%%%%%%%%%%%%%%%%%%%%%%%%%%%%%%%%%%%%%%%%%%%%%%%%

\section{Mass modifications---$m_{3/2}\lesssim\Lambda$}
\label{sec:almost}

In this section we discuss the potential mass modifications.  In Section
\ref{subsec:almost mat} we present the mass matrices and kinetic terms for the fermions,
and the gauge kinetic terms. In Section \ref{subsec:almost EW} we discuss the
constraint from the electroweak
$S$ parameter. In Sections \ref{subsec:almost qual} and \ref{subsec:almost num}
we carry out a qualitative analysis of the spectrum---first using an
approximation of large $\mu$, and then, since $\mu$ actually need
not be that large, a numerical evaluation of the spectrum.

%%%%%%%%%%%%%%%%%%%%%%%%%%%%%%%%%%%%%%%%%%%

\subsection{Mass matrices and kinetic terms}
\label{subsec:almost mat}

The dominant non-renormalizable contributions to the
Lagrangian terms of Section \ref{notation} yield a
neutralino mass matrix of the form
\begin{eqnarray}
  \begin{split}
    M_{\tilde{N}}&=\scriptsize\left(\begin{array}{cccc}M_1+\frac{ m_{3/2} v_u v_d }{\Lambda^2} & \frac{ m_{3/2} v_u v_d }{\Lambda^2} & A_1+ \frac{m_{3/2}^2 v_d }{\Lambda^2} & A_2+ \frac{m_{3/2}^2 v_u }{\Lambda^2} \\ \frac{ m_{3/2} v_u v_d }{\Lambda^2} & M_2+\frac{m_{3/2} v_u v_d }{\Lambda^2} &A_3+ \frac{m_{3/2}^2 v_d }{\Lambda^2} & A_4+ \frac{m_{3/2}^2 v_u }{\Lambda^2} \\
    A_1+ \frac{m_{3/2}^2 v_d }{\Lambda^2} & A_3+ \frac{m_{3/2}^2 v_d }{\Lambda^2} &  \frac{v_{u}^2}{\Lambda} & -\mu+\max(\frac{v_u v_d}{\Lambda},\frac{m_{3/2}^2 v_u^2}{\Lambda^3}) \\ A_2+ \frac{m_{3/2}^2 v_u }{\Lambda^2} & A_4+ \frac{m_{3/2}^2 v_u }{\Lambda^2} & -\mu+\max(\frac{v_u v_d}{\Lambda},\frac{m_{3/2}^2 v_u^2}{\Lambda^3}) &\max(\frac{v_{d}^2}{\Lambda}, \frac{v_u^2 m_{3/2}}{\Lambda^2}) \end{array}\right).
      \end{split}
      \label{mn m32 small}
\end{eqnarray}
In the chargino sector we have
\begin{eqnarray}
  \begin{split}
    M_{\tilde{C}}&=\left(\begin{array}{cc}
M_2+\frac{m_{3/2} v_u v_d}{\Lambda^2} &
-\sqrt{2}(A_4+\frac{m_{3/2}^2 v_u}{\Lambda^2}) \\
\sqrt{2}(A_3+\frac{m_{3/2}^2 v_d}{\Lambda^2}) & \mu-\max(\frac{v_u
v_d}{\Lambda},\frac{m_{3/2}^2 v_u^2}{\Lambda^3})
\end{array} \right)\;.
  \end{split}\label{mc m32 small}
\end{eqnarray}
The gauge coupling matrix becomes
\begin{eqnarray}
  \begin{split}
    {\cal{S}}= \left(\begin{array}
      {cc} -\frac{1}{4}+\max (\frac{v_u v_d}{\Lambda^2},\frac{m_{3/2}^2 v_u^2}{\Lambda^4}) & \max (\frac{v_u v_d}{\Lambda^2},\frac{m_{3/2}^2 v_u^2}{\Lambda^4}) \\ \max(\frac{v_u v_d}{\Lambda^2},\frac{m_{3/2}^2 v_u^2}{\Lambda^4}) & -\frac{1}{4}+\max (\frac{v_u v_d}{\Lambda^2},\frac{m_{3/2}^2 v_u^2}{\Lambda^4})
    \end{array}\right)\;,
    \label{S matrix m32 small}
  \end{split}
\end{eqnarray}
%\begin{eqnarray}
%  \begin{split}
%    {\cal{S}}= \left(\begin{array}
%      {cc} -\frac{1}{4}+\frac{m_{3/2}^2}{\Lambda^2} & \max (\frac{v_u v_d}{\Lambda^2},\frac{m_{3/2}^2 v_u^2}{\Lambda^4}) \\ \max(\frac{v_u v_d}{\Lambda^2},\frac{m_{3/2}^2 v_u^2}{\Lambda^4}) & -\frac{1}{4}+\frac{m_{3/2}^2}{\Lambda^2}
%    \end{array}\right)\;,
%    \label{S matrix m32 small}
%  \end{split}
%\end{eqnarray}
and the kinetic fermion mixing matrix can be written as
\begin{eqnarray}
  \begin{split}
      K&=\left( \begin{array}
      {cccc} \max(\frac{v_u v_d}{\Lambda^2},\frac{m_{3/2}^2 v_u^2}{\Lambda^4}) & \max(\frac{v_u v_d}{\Lambda^2},\frac{m_{3/2}^2 v_u^2}{\Lambda^4}) & \frac{m_{3/2} v_d}{\Lambda^2} & \frac{m_{3/2} v_u}{\Lambda^2}\\
             \max(\frac{v_u v_d}{\Lambda^2},\frac{m_{3/2}^2 v_u^2}{\Lambda^4}) & \max(\frac{v_u v_d}{\Lambda^2},\frac{m_{3/2}^2 v_u^2}{\Lambda^4}) & \frac{m_{3/2} v_d}{\Lambda^2} & \frac{m_{3/2} v_u}{\Lambda^2}\\
             \frac{m_{3/2} v_d}{\Lambda^2} & \frac{m_{3/2} v_d}{\Lambda^2} & \frac{v_u^2}{\Lambda^2} & \frac{v_u^2}{\Lambda^2}\\
             \frac{m_{3/2} v_u}{\Lambda^2} & \frac{m_{3/2} v_u}{\Lambda^2} & \frac{v_u^2}{\Lambda^2} & \frac{v_u^2}{\Lambda^2}
    \end{array}\right)+\mathbbm{1}\;.%=\\
  \end{split}
  \label{K_m32_small}
\end{eqnarray}

In the above, all higher-dimensional contributions should be
understood as being accompanied by ${\cal{O}}(1)$ coefficients.

These orders of magnitude are dictated by the lowest possible power of $m_{3/2}/\Lambda$ for each term, subject to deviations from this due to
(physical values of) $\tan \beta$.  For instance, the corrections to the $S$ parameter are proportional to $v^2$, as dictated by
the gauge quantum numbers. Hence, the leading corrections to the
off-diagonal terms of ${\cal S}$ are either $v_u v_d/\Lambda^2$, arising
from terms of the form $\int d^2\theta WBHH/ \Lambda^2$, or $v_u^2 m_{3/2}^2/\Lambda^4$, arising from
terms of the form $\int d^4\theta DWDBHH/ \Lambda^4$. Any other
term originating from higher-dimensional operators will suppress
these by a positive power of $v^2/\Lambda^2$, $\mu/\Lambda$ or
$m_{3/2}/\Lambda$, which are all smaller than one.  With this example in mind, details regarding the rest of the operators are relegated to Appendix \ref{app:almost origins}.

Let us briefly address the issue of higher-dimensional operators
involving only the Higgs sector.  Such operators can affect the Higgs scalar potential and spectrum (see, e.g., \cite{Brignole:2003cm} and more recently \cite{Dine:2007xi}) and the electroweak phase transition \cite{Blum:2008ym}.  They will be constrained by naturalness considerations and possibly by EDMs, and further constraints from upcoming experiments.  Assuming the $\mu$ problem has been solved, the discussed non-renormalizable contributions to
Higgsino mass terms are non-dominant in our analysis, and so we do not expect such constraints to affect the discussed qualitative behavior.

%%%%%%%%%%%%%%%%%%%%%%%%%%%%%%%%%%%

\subsection{Electroweak constraints}
\label{subsec:almost EW}

A requirement we impose is that the scales $\Lambda$ and $m_{3/2}$ are much larger than the electroweak
scale, i.e.,
\begin{eqnarray}
  v_d \lsim v_u\ll\Lambda, m_{3/2}\;.
\end{eqnarray}
For simplicity, we also assume
\begin{eqnarray}
  m_{3/2}\;,\;\Lambda>\mu>v\ \ \ \mathrm{and}\ \ \ m_{3/2}>\mu \tan \beta\;,
\end{eqnarray}
though the analysis can easily be generalized when these relations are relaxed.

Our goal is to evaluate the character and mass of the LSP and NLSP as $m_{3/2}/\Lambda$ varies, under the two constraints that the LSP is neutral and that electroweak bounds are not violated.  The latter of the two can already be addressed---the constraint
(\ref{S ref}) coming from the electroweak sector obtained from (\ref{S matrix m32 small}) implies that the stronger of the following must hold:
\begin{eqnarray}
\begin{split}
  &\frac{v_d^2}{\Lambda^2}\tan\beta < -S\times 10^{-3}\;,\ \ \
  \frac{m_{3/2}^2 v_d^2}{\Lambda^4} \tan^2 \beta<-S\times 10^{-3}\ \ .
  \end{split}
  \end{eqnarray}
Combining the two, we obtain the electroweak bound of
\begin{eqnarray}\label{EW m32 small}
  \Lambda>\max\left(v_d \sqrt{\frac{\tan \beta}{-S\times 10^{-3}}}\ ,\ \left(\frac{m_{3/2}^2 v_d^2 \tan ^2 \beta}{-S \times 10^{-3}}\right)^{1/4}\right)\;.
\end{eqnarray}
However, for the range of $m_{3/2}$ of interest (avoiding a very light LSP), taking $S$ around it's central value \eqref{ST values}, this bound actually plays no role\footnote{In practice, $m_{3/2}$ larger than $\sim16$~$\mathrm{\ TeV}$ suffices for this.} in restricting the scale $\Lambda$, which is already above $m_{3/2}$.

%%%%%%%%%%%%%%%%%%%%%%%%%%%%%%%%

\subsection{Qualitative analysis}
\label{subsec:almost qual}

As mentioned earlier, we find that in most of this regime of
$m_{3/2}\lesssim\Lambda$, the lightest neutralino and chargino are predominantly
Wino-like, composing the LSP and NLSP, which is similar to the
standard anomaly mediation scenario at leading order. Beyond leading
order, we find tree level mass splitting between the two of up to a few GeV (and even higher), much larger than the often-quoted (loop) number of
170~$\mathrm{\ MeV}$. Next we will show how to obtain this result both
analytically (in the limit of large $\mu$) and numerically (for a
broad range of parameters).

%%%%%%%%%%%%%%%%%%%%%%%%%%%%

\subsubsection{Leading order}
\label{sssec:almost LO}

First, we note that all higher-dimensional contributions to the
kinetic mixing matrix $K$ of (\ref{K_m32_small}) are much smaller
than unity. Therefore, as far as the kinetic term is concerned, the
usual degrees of freedom of Winos, Bino and Higgsinos can be used.
In fact, we will be neglecting the corrections to the kinetic mixing
terms that originate from higher-dimensional operators. Including
them does not change the conclusions qualitatively.

The analysis proceeds by examining the relative size of the higher-dimensional operators versus the MSSM and usual AMSB contributions of the neutralino mass matrix \eqref{mn m32 small} in each
of the diagonal and off-diagonal blocks separately, under the constraint
(\ref{EW m32 small}).  This will allow us to neglect some of the terms in each
entry and will simplify the comparison between the blocks, which we
do next.

A straightforward analysis within each block separately shows that:
\begin{itemize}
\item In the diagonal blocks, the MSSM-usual AMSB terms ($M_1,M_2,\mu$)
dominate over the contribution of the higher-dimensional operators. We
therefore neglect the higher-dimensional operators in these two blocks.
\item In the off-diagonal block, the ratio between the contributions can vary throughout the $m_{3/2}<
\Lambda$
range. The MSSM contributions are encoded in the $A_i$, and the
contribution of the higher-dimensional operators is characterized by
the scale
\begin{equation}
M \equiv \frac{m_{3/2}^2 v_u}{\Lambda^2}
\label{epsilon}\;.
\end{equation}

\end{itemize}
We see that the contribution of higher-dimensional
operators can be meaningful if the off-diagonal block is large enough. Hence,
we turn our attention to comparison between the blocks.

The upper-diagonal block is characterized by the scale $M_i\sim
10^{-3}m_{3/2}$, the lower-diagonal block by the scale $|\mu|$, and
the off-diagonal block by the scales $A_i$ and $M$. In
standard AMSB, one assumes $3 \lesssim |\mu|/M_2 \lesssim 6$
\cite{Gherghetta:1999sw} or at least $M_2< |\mu|$, resulting in a
Wino-like LSP rather than a Higgsino-like one.  We will assume the latter
holds, implying a weak constraint of $m_{3/2,max} \sim 10^3 |\mu|$,
which can easily be met. In the remainder of the work, we assume
this holds.  It is easy to see that the lower-diagonal block
dominates in this range over both the upper-diagonal and
off-diagonal blocks.  A similar analysis holds in the chargino sector for the mass matrix of \eqref{mc m32 small}.

Since the lower-diagonal block overshadows the others, the lightest
neutralino and chargino are predominantly Wino-like, composing the
LSP and NLSP. At leading order, the behavior is thus similar to the
standard AMSB scenario, and indeed the decoupling limit enters this region above a certain $\Lambda$. However, the
splitting in the Wino multiplet can be different from the usual
case, if the off-diagonal block is large enough. We turn next to this effect.

%%%%%%%%%%%%%%%%%%%%%%%%%%%%%%%%%%%%

\subsubsection{Next to leading order}
\label{sssec:almost NLO}

Beyond leading order, the non-renormalizable contributions to
neutralino and chargino masses \eqref{mn m32 small} and \eqref{mc m32 small} in the off-diagonal block introduce the largest
mass splitting in the triplet.\footnote{We can also consider another
tree level contribution, coming from the off-diagonal terms within the
upper-diagonal block, denoted by $\rho$.  The neutralino gets a mass shift of
\begin{eqnarray}
  \begin{split}
    -\frac{\rho^2}{M_1-M_2}=-\frac{m_{3/2} v_d^4 \tan ^2 \beta} {6.2 \times10^{-3}\Lambda^4}\;,\nonumber
  \end{split}
\end{eqnarray}
while the chargino has no such contribution.  This splitting,
however, is very small, and reaches a maximum splitting in parameter
space of $\sim $40 MeV for the range of $m_{3/2}$ of interest.}  In the limit of large $|\mu|$, the lightest neutralino
and chargino
masses are well approximated by
%%%%%%   taken from file "almost use"   %%%%%%
\begin{eqnarray}
\begin{split}
  &M_{\tilde{N}_1} \approx  M_2+\frac{2cd v_d^2}{\mu}-\frac{(ad+cb)^2 v_d^4}{\mu^2(M_1-M_2)}-\frac{(c^2+d^2)M_2v_d^2}{\mu^2}+\frac{2cdM_2^2 v_d^2}{\mu^3}\\
  &\ -\frac{2cd(c^2+d^2)v_d^4}{\mu^3}+\frac{2(ad+cb)(ac+bd)M_2v_d^4}{\mu^3(M_1-M_2)}+\frac{2(ad+cb)^2(a b-cd)v_d^6}{(M_1-M_2)^2\mu^3}\;,
  \end{split}
  \label{mn_standard_approx}
\end{eqnarray}

\begin{eqnarray}
  M_{\tilde{C}_1}\approx M_2+\frac{2cd v_d^2}{\mu}-\frac{(c^2+d^2)v_d^2 M_2}{\mu^2}+\frac{2cdM_2^2 v_d^2}{\mu^3}-\frac{2cd(c^2+d^2)v_d^4}{\mu^3}\;,
  \label{mc_standard_approx}
\end{eqnarray}
where we have denoted the off-diagonal block by
\begin{eqnarray}
  v_d\left(\begin{array}{cc} a & b \\ c & d \end{array}\right)=v_d\left(\begin{array}{cc} \tilde{a}\frac{m_{3/2}^2}{\Lambda^2}-\frac{g'}{\sqrt 2} & \tilde{b}\frac{m_{3/2}^2}{\Lambda^2}\tan\beta+\frac{g' \tan\beta}{\sqrt 2} \\ \tilde{c}\frac{m_{3/2}^2}{\Lambda^2}+\frac{g}{\sqrt2} & \tilde{d}\frac{m_{3/2}^2}{\Lambda^2}\tan\beta-\frac{g \tan \beta}{\sqrt2} \end{array}\right),
  \label{OD almost def}
\end{eqnarray}
and $\tilde{a},\tilde{b},\tilde{c},\tilde{d}$ are the ${\cal{O}}(1)$ coefficients in the
appropriate entries.  For simplicity, we consider the case of real ${\cal O}(1)$ coefficients here.  The mass splitting is then
%%%% taken from file "almost use"  %%%%
\begin{eqnarray}
  \begin{split}
    \Delta m \approx  v_d^4\left(\frac{(ad+bc)^2}{\mu^2(M_1-M_2)}-\frac{2(ad+cb)(ac+bd)M_2}{\mu^3 (M_1-M_2)}-\frac{2(ad+cb)^2(a b-cd)v_d^2}{(M_1-M_2)^2\mu^3}  \right)\nonumber.
  \end{split}\\
  \label{mass splitting standard}
\end{eqnarray}

At leading order in $1/\mu$, the mass splitting is always positive,
and so the LSP is always neutral.  However, we see that the
next-to-leading order mass splitting may be negative, and so the
${\cal{O}}(1)$ coefficients will have to be constrained. We note that the splitting of the triplet is indeed at order $v^4$ as group theory demands. The splitting has a weak dependence on the sign of $\mu$ and does not
depend on it at leading order in $1/\mu$.

The scale of splitting suggested by this argument, for example for $m_{3/2}/\Lambda\sim1$, $m_{3/2}\sim50$ TeV, $\mu$ of order a few hundred GeV and $\tan \beta=2$, is of order GeV. However, this is only a rough estimate
since $\mu$ is actually not large enough for the approximation above
to be correct. Hence, some numerical results are given in Table \ref{tab:num}.

%%%%%%%%%%%%%%%%%%%%%%%%%%%%%%

\subsection{Numerical results}
\label{subsec:almost num}

%%%%%%%%   ALL RESULTS ARE FROM "LARGE TAN BETA -> SHORTER -> CORRECTED", THE BLAH_BLAH_MOD.MAT FILES  %%%%%%

\begin{table}
\caption{$\tan \beta=2$ (top) and $\tan \beta=10$ (bottom)}
\label{tab:num}

\begin{center}
\begin{tabular}{|c |c | c | c |c | c |c |c |} \hline\hline
%\rule{0pt}{1.2em}%

\multicolumn{2}{|c|}{} &\multicolumn{3}{|c|}{$m_{3/2}/\Lambda=1$}&\multicolumn{3}{|c|}{$m_{3/2}/\Lambda=1/2$}\\ \hline \hline

$\mu$ &  $m_{3/2}\ (m_{AMSB})$ & $m_{LSP}$ &  $\Delta m_{av}$  & ${\Delta m}_{10}$  & $m_{LSP}$ & $\Delta m_{av}$ & ${\Delta m}_{10}$  \\

\scriptsize[GeV] & \scriptsize[TeV] (\scriptsize[GeV]) & \scriptsize[GeV] & \scriptsize[GeV] & \scriptsize[GeV] & \scriptsize[GeV] &  \scriptsize[GeV] & \scriptsize[GeV] \\
\hline\hline

%mu=400
\multirow{2}{*}{$400$}
 & $50\ (116)$ & $110\pm 34$ & $2.3$ & $6.5$ & $113\pm23$ &
$1.2$ & $3.4$ \\
 & $100\ (235)$ & $225\pm 46$ & $3.4$ & $10$ & $231\pm34$ & $2$ &
$5.2$ \\ \hline

%mu=800
\multirow{2}{*}{$800$}
& $50\ (127)$ & $126\pm 17$ & $0.5$ & $1.4$ & $127\pm11$ &
$0.2$ & $0.6$ \\

 & $100\ (260)$ & $257\pm 20$ & $0.4$ & $0.8$ & $258\pm13$ & $0.2$ & $0.3$ \\ \hline\hline

\end{tabular}
\end{center}

\begin{center}
\begin{tabular}{|c |c | c | c |c | c |c |c |} \hline\hline
%\rule{0pt}{1.2em}%

\multicolumn{2}{|c|}{} &\multicolumn{3}{|c|}{$m_{3/2}/\Lambda=1$}&\multicolumn{3}{|c|}{$m_{3/2}/\Lambda=1/2$}\\ \hline \hline

$\mu$ &  $m_{3/2}\ (m_{AMSB})$ & $m_{LSP}$ &  $\Delta m_{av}$  & ${\Delta m}_{10}$  & $m_{LSP}$ & $\Delta m_{av}$ & ${\Delta m}_{10}$  \\

\scriptsize[GeV] & \scriptsize[TeV] (\scriptsize[GeV]) & \scriptsize[GeV] & \scriptsize[GeV] & \scriptsize[GeV] & \scriptsize[GeV] &  \scriptsize[GeV] & \scriptsize[GeV] \\
\hline\hline

%mu=400
\multirow{2}{*}{$400$}
& $50\ (126)$ & $119\pm 17$ & $1.3$ & $3.8$ & $122\pm12$ &
$0.7$ & $2$ \\
 & $100\ (248)$ & $235\pm 33$ & $3$ & $8.4$ & $242\pm26$ & $2$ &
$4.9$ \\ \hline

%mu=800
\multirow{2}{*}{$800$}
& $50\ (132)$ & $130\pm 7$ & $0.16$ & $0.4$ & $131\pm5$ &
$0.08$ & $0.14$ \\

 & $100\ (265)$ & $260\pm 11$ & $0.3$ & $1$ & $263\pm8$ & $0.15$ &
$0.3$ \\ \hline\hline

\end{tabular}
\end{center}

\end{table}
In Table \ref{tab:num} we present numeric evaluation of the masses for $m_{3/2}=50,\ 100$~$\mathrm{\ TeV}$, $\Lambda=(1,2)m_{3/2}$, $\mu=400,\ 800$~$\mathrm{\ GeV}$ and $\tan\beta=2,\ 10$ (negative $\mu$ values give rise to very similar results).  These are the input parameters.  For each one we evaluate%(negative $\mu=-400,\ -800$ GeV give rise to very similar results).  These are the input parameters. For each one we evaluate
\begin{itemize}
 \item$m_{AMSB}$ - the LSP mass in AMSB without higher-dimensional terms
 \item $m_{LSP}$ - the average LSP mass (which is roughly similar to $m_{AMSB})$ and its standard deviation
 \item $\Delta m_{av}$ - the average mass difference to the NLSP
 \item $\Delta m_{10}$  - the mass difference to the NLSP above which $10\%$ of the mass splittings occur.
\end{itemize}
$m_{3/2}$ is given in TeV and the rest in GeV. The LSP is almost always the lightest neutralino, and the NLSP the lightest chargino.\footnote{The reverse case happens in a few percent of the runs and we drop those from the statistics.}
The numerical calculation was carried out using random ${\cal{O}}(1)$ coefficients for the various terms in the matrices in Section \ref{subsec:almost mat}, obtained from a uniform distribution $U[-1,1]$, where we scaled the coefficients such that the largest in the neutralino mass matrix is $\pm1$. The maximal mass splitting is indeed of order a few GeV.

Comparing $\Delta m_{av}$ and $\Delta m_{10}$, one notes that the distribution has a long tail. Indeed, much higher values of the mass splitting can be reached---for example at the level of $1\%$ of the simulations, one can reach splitting of order $20-25,\ 4-14$~$\mathrm{\ GeV}$ for $\tan \beta=2,\ 10$ at $\mu=400$~$\mathrm{\ GeV}$, or $8-12,\ 1.5-2$~$\mathrm{\ GeV}$ at $\mu=800$~$\mathrm{\ GeV}$.

The maximal reach of the mass splitting for a given $\mu$ decreases as $\tan \beta$ increases.
Also, as expected, the large mass splittings occur for $m_{3/2}\sim\Lambda$, where the LSP and NLSP
degrees of freedom begin mixing and are no longer pure Winos, but typically
are still predominantly so and form a triplet.  The average mass splitting becomes of the same order as the standard AMSB value $\sim 170$ MeV at $m_{3/2}/\Lambda\sim 0.4-0.8$ for various values of $m_{3/2}$, $\tan \beta$ and $\mu$.

%%%%%%%%%%%%%%%%%%%%%%%%%%%%%%%%%%%%%%%%%%%%%%%%

\section*{Acknowledgments}

We would like to thank Y. E. Antebi, K. Blum and Y. Nir, and in particular Z. Komargodski and Y. Shadmi, for illuminating discussions and insights.  This work was supported by the Israel-U.S. Binational Science Foundation, by a center of excellence supported by the Israel Science Foundation (grant number
1468/06), by a grant (DIP H52) of the German Israel Project Cooperation, by the
European network MRTNCT-2004-512194, by a grant from G.I.F., the German-Israeli Foundation for Scientific Research and Development, and by the Einstein-Minerva center for theoretical physics.

\begin{appendix}

%%%%%%%%%%%%%%%%%%%%%%%%%%%%%%%%%%%%%%%%%%%%%%%%%%%%%%%%%%

\section{Some higher powers of $m_{3/2}/\Lambda$}
\label{app:higher}

In this appendix we evaluate the role of higher powers of $m_{3/2}/\Lambda$, which may become important as this ratio approaches unity. We will show that even when including such terms, the qualitative results do not change. We consider contributions of higher powers in $m_{3/2}/\Lambda$ to the
terms of Section \ref{notation} that are still in the Lagrangian \eqref{lagrule}.  This yields contributions to the following terms:

\begin{itemize}
\item In the neutralino mass matrix, the upper-diagonal block has contributions of the form $m_{3/2}^2 v_u v_d/\Lambda^3$, to be compared with the previous $M_i$ and $m_{3/2} v_u v_d/\Lambda^2$.  The off-diagonal block already contains the highest power of $m_{3/2}$ and so is unchanged.  In the lower-diagonal block, the $\tilde{H}_d \tilde{H_d}$ term has contributions of the form $m_{3/2}^2 v_u^2/\Lambda^3$, to be compared with the previous $v_u^2/\Lambda$. The chargino mass matrix is altered accordingly.

\item The gauge coupling matrix ${\cal{S}}$ remains the same.
\item The kinetic mixing matrix $K$ has contributions in the off-diagonal blocks, mixing Wino/Bino-Higgsino kinetic terms.  The contribution in each of these entries is as was previously, altered by an additional factor of $m_{3/2}/\Lambda$.
\end{itemize}
All the above contributions are thus similar to previous terms, with additional factors of $m_{3/2}/\Lambda$ which is approaching one.  This is not very surprising since when $m_{3/2}=\Lambda$ the strength of all higher-dimensional contributions is determined by powers of $v_u/\Lambda$ or $v_d/\Lambda$, which for the most part are dictated by SM quantum numbers.  %It is easily seen that including these new terms, the upper-and lower-diagonal blocks are still dominated by the renormalizable terms, while in the off-diagonal block the same non-renormalizable contributions as before have the upper hand.
This implies that the inclusion of such terms as $m_{3/2}$ approaches $\Lambda$ does not change the qualitative behavior we found; and indeed numerical  results support this.

%%%%%%%%%%%%%%%%%%%%%%%%%%%%%%%%%%%%%%%%%%%%%%%%%%%%%%%%%%

\section{Origins of orders of magnitude}
\label{app:almost origins}

In this appendix we outline the dominant contributions to the various Lagrangian terms of Section \ref{subsec:almost mat}.  These arise from the following (appropriate chiral and anti-chiral superfields are to be understood in the context, as well as contraction of gauge and Lorentz indices):

\begin{itemize}

\item The leading corrections to the matrix ${\cal{S}}$ subject to $\tan\beta$ are either $v_u v_d/\Lambda^2$, arising, e.g., from terms of the form $\int d^2\theta G G'HH/ \Lambda^2$, or $v_u^2 m_{3/2}^2/\Lambda^4$, arising, e.g., from
terms of the form $\int d^4\theta DGDG'HH/ \Lambda^4$, where $G$ and $G'$ are each either $W$ or $B$.

%\item The corrections to the $S$ parameter are proportional to $v^2$ by gauge quantum numbers.  Thus, as already mentioned, the leading corrections to the off-diagonal terms of ${\cal S}$ subject to $\tan \beta$ are either $v_u v_d/\Lambda^2$, arising, e.g., from terms of the form $\int d^2\theta WBHH/ \Lambda^2$, or $v_u^2 m_{3/2}^2/\Lambda^4$, arising, e.g., from
%terms of the form $\int d^4\theta DWDBHH/ \Lambda^4$.

%The leading corrections to the gauge kinetic terms of $W$ and $B$ separately in
%${\cal{S}}$ are $m_{3/2}^2/\Lambda^2$, arising for instance from terms of the
%form $\int d^4 \theta DGDG/\Lambda^2$, where $G$ is either $W$ or
%$B$.  This amounts to rescaling the weak and hypercharge gauge
%couplings.

\item The leading corrections to direct Wino-Bino mass mixing are
of order $v^2 m_{3/2}/\Lambda^2$, which come from operators such as
$\int d^2 \theta GG'HH/\Lambda^2$, where $G$ and $G'$ are each
either $W$ or $B$. %The factor of $v^2$ is determined by the
%gauge-group structure.

\item In the Wino/Bino-Higgsino mass mixing, the SM contribution is
proportional to $v$. The leading correction is of order
$vm_{3/2}^2/\Lambda^2$, generated, for instance, by an operator
$\int d^4 \theta G D H H/ \Lambda^2$, where $G$ is either $W$ or
$B$.

\item Higgsino-Higgsino mass corrections can be of order $v^2/
\Lambda$, $v_u^2 m_{3/2}/\Lambda^2$, or $v_u^2 m_{3/2}^2/\Lambda^3$, e.g., from operators of the sort $\int d^2 \theta
HHHH/ \Lambda$, $\int d^4 \theta HHHH/\Lambda^2$ or $\int d^4 \theta DHDH HH/\Lambda^3$, respectively.  These are the leading contributions for the various entries allowed by the gauge structure and when taking $\tan \beta$ effects into account.  We do not allow terms of the form $\int d^4 \theta
H_u H_d$ which, after inserting the compensator, will give rise to a $\mu$-term of order $m_{3/2}$.

\item %The leading corrections to Wino-Wino and Bino-Bino kinetic
%mixing are of order $m_{3/2}^2/\Lambda^2$, arising for instance from terms of the
%form $\int d^4 \theta DGDG/\Lambda^2$, where $G$ is either $W$ or
%$B$.
%
The leading corrections to the kinetic Wino-Bino mixing subject to $\tan\beta$ are of order $v^2/\Lambda^2$ or $v_u^2 m_{3/2}^2/\Lambda^4$, arising for instance from $\int d^2 \theta GG'HH/ \Lambda^2$ or $\int d^4 \theta G D^2 G' H H/\Lambda^4$, respectively, where $G$ and $G'$ are each either $W$ or $B$.

\item In the Bino/Wino-Higgsino kinetic mixing, the leading
contributions are $m_{3/2} v/\Lambda^2$.  Such terms come for
instance from $\int d^4 \theta
DGHH/ \Lambda^2$ or $\int d^4 \theta GDHH/\Lambda^2$, where $G$ is $W$ or $B$.

\item The leading corrections to kinetic mixing among the Higgsinos are
also the maximal allowed by the gauge structure, i.e.,
$v^2/\Lambda^2$, arising for instance from terms of the form $\int
d^4 \theta HHHH/ \Lambda^2$.

\end{itemize}

\end{appendix}

%%%%%%%%%%%%%%%%%%%%%%%%%%%%%%%%%%%%%%%%%%%%%%%%%%%%%%%%%%%%%%%%%%%%%%%%

\end{document}